\documentclass[a4paper,10pt]{article}
\usepackage{graphicx}
\usepackage{comment}
\usepackage{amssymb}
\usepackage{amsmath}
\usepackage{nicefrac}
\usepackage{graphicx}
\usepackage{dcolumn}
\usepackage{bm}
\usepackage{comment}
\usepackage{multirow}

\begin{document}

\huge

\begin{center}
A note on the contribution of multi-photon processes to radiative opacity
\end{center}

\vspace{0.5cm}

\large

\begin{center}
Jean-Christophe Pain\footnote{jean-christophe.pain@cea.fr}
\end{center}

\normalsize

\begin{center}
\it CEA, DAM, DIF, F-91297 Arpajon, France
\end{center}

\vspace{0.5cm}

\begin{abstract}
Recently, Bailey \emph{et al.} performed iron opacity measurements on the Z machine at Sandia National Laboratory in conditions close to the ones of the base of the convective zone of the Sun. Such experiments have raised questions about the physical models commonly used in opacity codes. To understand the discrepancy between experiment and theory, More \emph{et al.} investigated the role of two-photon processes. In the present work we show, by a simple estimate and using hydrogenic formulas, that due to the intensity of the backlight radiation seen by the sample, such processes are likely to play an important role only for highly excited states. 
\end{abstract}

\section{Introduction}

Two-photon absorption opacity was calculated by More and Rose in 1991 \cite{more91} using a semi-classical method. Recently, More \emph{et al.} presented a tentative study of such processes \cite{more17} in order to quantify their contribution to the opacity in the conditions of the recent opacity measurement on the Z machine by Bailey \emph{et al.} at $T$=182 eV and $n_e$=3.1$\times$10$^{22}$ cm$^{-3}$, conditions close to the ones of the frontier between the radiative and convective zones of the Sun \cite{bailey15}. In the latter experiment, the inferred opacity was found to be 30 to 400 \% higher than all the calculations, which represents a puzzling enigma for theorists. The preliminary numerical calculations published in Ref. \cite{more17} give substantial cross-sections comparable to the extra opacity observed in experiments on the Sandia Z-machine, but without yielding agreement with the experimental data. 

We also believe that multi-photon absorption deserves scrutiny. In the present note, we show that the radiation from the backlighter source, due to its brightness temperature, ensures that electric-dipole two-photon processes are of the same order of magnitude as one-photon ones for highly excited states, which are not occupied due to the relatively high density ($n_e$=3.1$\times$10$^{22}$ cm$^{-3}$, corresponding to $\rho\approx$ 0.17 g/cm$^3$).

The multi-photon absorption can be calculated by second-order perturbation theory as a sum over intermediate states with dipole matrix elements linking the initial state to various excited states \cite{gontier71}. The radial integral of dipolar matrix element between $n\ell$ and $n'\ell'$ subshells is

\begin{equation}
\mathcal{R}_{n\ell}^{n'\ell'}=\int_0^{\infty}R_{n\ell}(r)R_{n'\ell'}(r)r^3dr,
\end{equation}

\noindent where $R_{n\ell}$ is the radial part of the wavefunction, solution of Schr\"odinger's equation

\begin{equation}
-\frac{1}{2r^2}\frac{d}{dr}\left[r^2\frac{dR_{n\ell}(r)}{dr}\right]+\frac{\ell(\ell+1)}{2r^2}R_{n\ell}(r)+V(r)R_{n\ell}(r)=E_{n\ell}R_{n\ell}(r),
\end{equation}

\noindent where $V(r)$ is the electrostatic potential and $E_{n\ell}$ the eigen-energy. For a two-photon transition, one has to replace the interaction potential

\begin{equation}
-eF\mathcal{R}_{n\ell}^{n'\ell'},
\end{equation}

\noindent where $e$ is the electron charge and $F$ the amplitude of the incident electric field, by \cite{pindzola78,bassani77}

\begin{equation}\label{crux}
-e^2F^2\sum_{n''\ell''}\mathcal{R}_{n\ell}^{n''\ell''}\frac{1}{E_{n\ell}-E_{n''\ell''}-\hbar\omega}\mathcal{R}_{n''\ell''}^{n'\ell'},
\end{equation}

\noindent where $n''\ell''$ are intermediate subshells \cite{bassani77}. The energy denominators are differences of initial and intermediate state energies, including the photon energies. It is important to mention that equation (\ref{crux}) reflects only the radial contribution to the two-photon process. Angular factors as well as polarization effects are considered to be of the order of unity and as such play a role in the complete calculation but do not affect the scaling argument presented in this article. In the Sandia experiment, the main transitions involving shells $n$=2, 3 and 4, it is reasonable to consider cases where $n'$ and $n$ are close to each other and intermediate states with energies close to the ones of $n\ell$ and/or $n'\ell'$, yielding the most important contribution to the matrix elements. For instance, one may have \cite{more17}\footnote{More \cite{more17} provides a good discussion on how two-photon processes may arise between initial and final configurations. Subtle details regarding Pauli blocking that may occur in the intermediate state are not discussed here and the interested reader should refer to the aforementioned work.}

\begin{equation}
1s^22s^22p^6\rightarrow \left\{\begin{array}{l}1s^22s2p^63p \\1s^22s^22p^53d\end{array}\rightarrow 1s^22s2p^53p3d.\right.
\end{equation}

In the hydrogenic approximation, one has

\begin{equation}
V(r)=-\frac{Z_{\mathrm{eff}}e^2}{4\pi\epsilon_0 r}
\end{equation}

\noindent where $Z_{\mathrm{eff}}$ represents the screened hydrogenic charge effectively seen by the electron. Neglecting the Lamb shift, the energy difference involved in the denominator of Eq. (\ref{crux}) behaves as (``$\mathrm{Ryd}$'' denotes the Rydberg constant):

\begin{eqnarray}\label{deltae}
E_{n\ell}-E_{(n+1) \ell''}&=&-\mathrm{Ryd}\frac{Z_{\mathrm{eff}}^2}{n^2}+\mathrm{Ryd}\frac{Z_{\mathrm{eff}}^2}{(n+1)^2}\nonumber\\
&=&-\mathrm{Ryd}\left(\frac{2n+1}{n^2(n+1)^2}\right)Z_{\mathrm{eff}}^2,
\end{eqnarray}

\noindent and varies as $1/n^3$. The dipole integrals are given by Gordon's formula \cite{gordon29}:

\begin{eqnarray}\label{gordon}
\mathcal{R}_{n\ell}^{n'(\ell-1)}&=&\frac{a_0}{4Z_{\mathrm{eff}}}\frac{(-1)^{n'-\ell}}{\left(2\ell-1\right)!}\sqrt{\frac{(n+\ell)!(n'+\ell-1)!}{(n-\ell-1)!(n'-\ell)!}}\frac{(4nn')^{\ell+1}(n-n')^{n+n'-2\ell-2}}{(n+n')^{n+n'}}\nonumber\\
& &\times\left\{_2F_1\left(-n_r,-n'_r,2\ell,-\frac{4nn'}{(n-n')^2}\right)\right.\nonumber\\
& &-\left.\left(\frac{n-n'}{n+n'}\right)^2~_2F_1\left(-n_r-2,-n'_r,2\ell,-\frac{4nn'}{(n-n')^2}\right)\right\},
\end{eqnarray}

\noindent where $_2F_1$ is the Gauss hypergeometric function, $a_0$ the Bohr radius, $n_r=n-\ell-1$ and $n'_r=n'-\ell$. 

When the initial and final states are in the same shell ($\Delta n=0$ transitions), the radial dipole matrix element reads

\begin{equation}\label{inside}
\mathcal{R}_{n\ell}^{n (\ell-1)}=\mathcal{R}_{n (\ell-1)}^{n\ell}=\frac{3a_0}{2Z_{\mathrm{eff}}}n\sqrt{n^2-\ell^2},
\end{equation}

\noindent which varies as $n^2$. Since in that case the energy difference is zero in the hydrogenic appproximation without fine structure, we consider transitions for which $n'\ne n$. In the case where only one of the two values $n$ or $n'$ is large (for instance $n$), we find from Eq. (\ref{gordon}) that the radial matrix element varies as $n^{-3/2}$. For instance, for the $2p\rightarrow nd$ transition, we have \cite{bethe57}:

\begin{equation}
\left(\mathcal{R}_{n2}^{21}\right)^2=\left(\frac{a_0}{Z_{\mathrm{eff}}}\right)^2\frac{2^{19}n^9\left(n^2-1\right)(n-2)^{2n-7}}{3(n+2)^{2n+7}}.
\end{equation}

This dependence can be easily understood, noticing that the radial part of the wavefunction can be put in the form \cite{shakeshaft79}:

\begin{equation}
R_{n\ell}(r)=A_{n\ell}\sum_{k=0}^{\infty}(-1)^ka_k\left(2\frac{Z_{\mathrm{eff}}r}{a_0}\right)^{\frac{k-1}{2}}J_{2\ell+1+k}\left(\sqrt{8\frac{Z_{\mathrm{eff}}r}{a_0}}\right),
\end{equation}

\noindent where $J_n$ is the Bessel function of order $n$ and 

\begin{equation}
a_k=\frac{1}{4kn^2}\left[(2\ell+k)a_{k-2}+a_{k-3}\right]\;\; \mathrm{for}\;\; k\ge 3,\;\;\;a_0=1,\;\;\;a_1=0,\;\;\;a_2=\frac{\ell+1}{4n^2},
\end{equation}

\noindent and

\begin{equation}
A_{n\ell}=2\left(\frac{Z_{\mathrm{eff}}}{a_0}\right)^3\frac{1}{n^{3/2}}\left[\prod_{k=1}^{\ell}\left(1-\frac{k^2}{n^2}\right)\right]^{1/2}.
\end{equation}

The first term of the series for small values of $r$ is 

\begin{equation}
R_{n\ell}(r)\approx A_{n\ell}\frac{1}{\sqrt{2\frac{Z_{\mathrm{eff}}r}{a_0}}}J_{2\ell+1}\left(\sqrt{8\frac{Z_{\mathrm{eff}}r}{a_0}}\right).
\end{equation}

Since $J_{2\ell+1}\left(\sqrt{8\frac{Z_{\mathrm{eff}}r}{a_0}}\right)/\sqrt{2\frac{Z_{\mathrm{eff}}r}{a_0}}$ is independent of $n$, $R_{n\ell}(r)$ behaves as $n^{-3/2}$. The matrix element $\mathcal{R}_{n\ell}^{n' (\ell-1)}$ is nothing else than the overlap integral of the wavefunction of the highly-excited state with $rR_{n' (\ell-1)}$, the latter differing notably from 0 only for small values of $r$ ($r<n^2a_0/Z_{\mathrm{eff}}$). $\mathcal{R}_{n\ell}^{n' (\ell-1)}$ is therefore only sensitive to the part of the Rydberg-state wavefunction $rR_{n\ell}$ close to the nucleus, where it only scales with $n$ as $n^{-3/2}$. In that case, the two-photon matrix element varies as 

\begin{equation}
\frac{n^{-3/2}n^{-3/2}}{n^{-3}}\approx 1.
\end{equation}

But, in the case where $n$ and $n'$ are both large, and close to each other (small values of $\Delta n$), the argument of the hypergeometric functions in Gordon's formula increases with the product $nn'$. The hypergeometric functions reduce then to derivative of Bessel functions \cite{abramowitz64}. It can be shown that, when $n_r,n'_r,n\rightarrow\infty$:

\begin{eqnarray}
& &~_2F_1\left(-n_r,-n'_r,2\ell;-4\frac{n(n+\Delta n)}{\Delta n^2}\right)\nonumber\\
& &\;\;\;\; \approx\frac{n'_r!(2\ell-1)!}{(n_r+2\ell-1)!}\left(-\frac{4n(n+\Delta n)}{\Delta n^2}\right)^{n_r}\left(\frac{\Delta n}{2}\right)^{-\Delta n}J'_{\Delta n}(\Delta n),
\end{eqnarray}

\noindent where $J'_n$ is the derivative of the Bessel function $J_n$. This yields

\begin{equation}
\mathcal{R}_{n\ell}^{n' (\ell\pm 1)}=\frac{3}{2}n'^2\frac{a_0}{Z_{\mathrm{eff}}}\left[\frac{2}{3\Delta n}J'_{\Delta n}(\Delta n)+O\left(\frac{\Delta n}{n'}\right)\right],
\end{equation}

\noindent which reveals a scaling with $n'$ or $n$ close to the one of Eq. (\ref{inside}) for transitions inside the same shell.

Therefore, while the one-electron matrix element (denoted $\mathcal{R}$ in the following) given by Eq. (\ref{gordon}) varies as $n^2$, the two-photon matrix element varies as $n^7$. In the same way, for a $N$-photon transition, the matrix element is the ratio of the product of $N$ dipolar elements by the product of $(N-1)$ energy differences. Its variation with respect to the principal quantum number is then:

\begin{equation}
\frac{\left(n^2\right)^N}{\left(n^{-3}\right)^{N-1}}=n^{5N-3}.
\end{equation}

Therefore, to a given intensity of the incident electric field corresponds a threshold value of $n$ above which the multi-photon processes dominate the linear effect, the threshold value decreasing with the incident flux. The single-photon rate is proportional to $2\pi\alpha$, and the two-photon rate to $4\pi^2\alpha^2$ \cite{bebb66}. In order to estimate such a threshold for two-photon processes, we write

\begin{equation}
eF\mathcal{R}\approx e^2F^2\frac{\mathcal{R}\times \mathcal{R}}{\Delta E}\times (2\pi\alpha).
\end{equation}

\noindent Using Eq. (\ref{deltae}), which can be approximated by

\begin{equation}
\Delta E\approx 2\mathrm{Ryd}\frac{Z_{\mathrm{eff}}^2}{n^3},
\end{equation}

\noindent we get

\begin{equation}
eF\mathcal{R}\approx\frac{2\mathrm{Ryd}Z_{\mathrm{eff}}^2}{2\pi\alpha n^3}.
\end{equation}

Inserting

\begin{equation}
\mathcal{R}\approx\frac{3}{2}\frac{a_0}{Z_{\mathrm{eff}}}n^2,
\end{equation}

\noindent we finally have 

\begin{equation}
F\approx\frac{Z_{\mathrm{eff}}^3e}{12\pi^2\alpha\epsilon_0a_0^2n^5}\approx Z_{\mathrm{eff}}^3\frac{7.5\times 10^{12}}{n^5}~\mathrm{V/m},
\end{equation}

\noindent where $\epsilon_0$ is the permittivity of vacuum. Assuming that, in the conditions of the experiment performed by Bailey \emph{et al.} \cite{bailey15}, we have  $Z_{\mathrm{eff}}\approx 15$, we get $F\approx 2.5\times10^{16}/n^5$ V/m. It means that the minimum flux required in order to ensure that two-photon processes start to be as important as one-photon processes is

\begin{equation}\label{phimin}
\Phi_{\mathrm{min}}=c\epsilon_0\frac{F^2}{2}\approx\frac{8.5\times 10^{25}}{n^{10}}~\mathrm{W/cm^2},
\end{equation}

\noindent where $c$ is the speed of light. In the conditions of the Z-pinch experiment, the backlighter (BL) can be considered as a Planckian distribution with a radiation temperature of $T_{\mathrm{BL}}\approx$ 350 eV. The dilution factor due to the geometry of the experiment is about 0.13 (see Refs. \cite{bailey15,nagayama14,bailey08}), which corresponds to an effective temperature of $T_{\mathrm{eff}}=\left(0.13\times T_{\mathrm{BL}}^4\right)^{1/4}\approx$ 210 eV. Therefore, the flux on the sample is about

\begin{equation}
\sigma T_{\mathrm{eff}}^4=0.13\times\sigma T_{\mathrm{BL}}^4\approx 2\times 10^{14}~\mathrm{W/cm^2},
\end{equation}

\noindent where $\sigma$ is Stefan's constant. This value is larger than $\Phi_{\mathrm{min}}$ (see Eq. (\ref{phimin})) for $n\geq 15$ (15$^{10}$=576,650,390,625). Therefore, for $n\approx$ 14-15, two-photon processes are as important as one photon-processes, but the former start to play a significant role for lower values of $n$, of course. However, in the conditions of the experiment ($T$=182 eV and $\rho\approx$ 0.17 g/cm$^3$), the highest populated shell predicted by the ion-sphere model corresponds to $n$=8 (truncation due to the density). Therefore, the contribution of two-photon processes should not explain the discrepancy between experiment and theory. 

Beyond the threshold $\Phi_{\mathrm{min}}$ (see Eq. \ref{phimin}), a perturbative calculation can not be used, since the expansion in powers of $F$ does not converge. In particular, the ionization probability is not negligible, and the absorption probability becomes continuous. A way to treat the phenomenon is then to use the semi-classical approach, in which the electromagnetic wave and the highly excited electron are treated simultaneously.

\section{Acknowledgments}

The author expresses his profound gratitude to the anonymous referee for valuable criticisms and suggestions.

\end{document}